\documentclass[11pt,a4paper]{article}
\usepackage{jheppub}

\usepackage{amsmath,amssymb,pxfonts}
\usepackage{verbatim,graphicx}

\usepackage{setspace}
\usepackage{color}
\usepackage[all]{xy}

\newcommand{\BR}{\mathbb{R}}

\newcommand{\beq}{\begin{equation}}
\newcommand{\beqs}{\begin{equation*}}
\newcommand{\eeq}{\end{equation}}
\newcommand{\eeqs}{\end{equation*}}

\allowdisplaybreaks
\begin{document}
\setlength{\unitlength}{1mm}
%%%%%%%%%%%%%%%%%%%%%%%%%%%%%%%%%%%%%%%%%%%%%%%%%%%%%%%%%%%%%%%%
\title{Open giant magnons suspended between dual giant gravitons in ${\cal N}=4$ SYM}

\author{David Berenstein,}
\author{Adolfo Holguin}
\affiliation { Department of Physics, University of California at Santa Barbara, CA 93106
}
\emailAdd{dberens@physics.ucsb.edu}
\emailAdd{adolfoholguin@physics.ucsb.edu}
\abstract{
We study classical solutions to the Nambu-Goto string on $AdS_2 \times S^1$ and $AdS_3 \times S^1$ corresponding to strings stretched between wrapped branes extending in $AdS$. The solutions are obtained by analytic continuation of giant magnon solutions, cut at the position of the branes.
 These solutions carry one or two $SO(2,4)$ charges and a single $SO(6)$ charge. We compute their energies and show their relation to $\frac{1}{2}$-BPS geometries. Their relevance to the $SL(2)$ sector of $\mathcal{N}=4$ SYM is also discussed.}

\maketitle

%%%%%%%%%%%%%%%%%%%%%%%%%%%%%%%%%%%%%%%%%%%%%%%%%%%%%%%%%%%%%%%%
\section{Introduction}
\label{S:Introduction}

The AdS/CFT correspondence has established a bridge between quantum gravity in asymptotically AdS spaces and gauge field theories \cite{Maldacena:1997re}. This connection makes it possible to extract information about strong coupling dynamics from calculations in the dual geometry. A particularly important example is the duality between string theory in $AdS_5\times S^5$ and ${\cal N}=4 $ SYM theory. In particular, the string excitations associated to the t'Hooft limit of the theory can be understood in a lot of detail from the
 study of planar diagrams in the field theory.

An important feature of planar $\mathcal{N}=4$ SYM is that it admits a description as an integrable spin-chain that computes the spectrum of anomalous dimensions of single trace operators (see \cite{Beisert:2010jr} for a textbook-level review). Relative to a ferromagentic ground state,
 it is known by supersymmetry arguments that in the infinite length limit of the $SU(2)$ sector of the spin-chain, single magnon excitations are characterized by the exact dispersion relation \cite{Santambrogio:2002sb,Beisert:2005tm}
\begin{equation}
\begin{aligned}
&\Delta- J = \sqrt{1+ \frac{\lambda}{\pi^2}\sin^2\left( \frac{p}{2}\right)}\\
\end{aligned}
\end{equation}
Here,  $J$ is a Cartan generator of the $SO(6)$ R-charge, $\lambda$ is the t' Hooft coupling and $p$ is the quasi-momentum of the magnon impurity.
The infinite chain limit corresponds to the limit $J\to \infty$, while $\Delta-J$ remains finite.
The dispersion relation follows from the symmetry algebra of the spin chain, and it is an all orders result in the t'Hooft coupling $\lambda$. The infinite spin chain has a centrally extended supersymmetry algebra \cite{Beisert:2005tm}, and the magnons are in short multiplets of the centrally extended algebra. As such, their contribution to the anomalous dimension is constrained by the kinematics of the shortening condition on the multiplet.
This shortness of the multiplet is a BPS condition for the states represented on the spin chain dynamics, but they are not global BPS states with respect to the global conformal field theory.

 The scaling of the dispersion relation at strong coupling suggests that these states should have simple descriptions in terms of classical strings. Such string solutions have been widely studied. They require a particular double scaling limit  that allows for the t' Hooft coupling to be a finite and adjustable parameter \cite{Hofman:2006xt}. This is in contrast to   the plane wave limit where $\lambda\to \infty$ while taking the long wavelength limit on the spin chain simultaneously, $\lambda/J^2 < \infty$ \cite{Berenstein:2002jq}. The Hoffman-Maldacena limit allows for a different reliable extrapolation of quantities between the strong and weak coupling regimes. Bound states of $Q$  giant magnons also have simple dispersion relations that have been computed both in the spin-chain model and in the classical string limit. Their contribution to the anomalous dimension also result from understanding short representations of the spin chain  algebra \cite{Chen:2006gea,Dorey:2006dq}; the dispersion relation of these bound states is given by
\begin{equation}
\begin{aligned}
\Delta- J &= \sqrt{Q^2+ \frac{\lambda}{\pi^2}\sin^2 \left(\frac{p}{2}\right)}\\
 J_2&= Q
\end{aligned}
\end{equation}
where $J_2$ is  a second Cartan generator of the $SO(6)$ symmetry (which is orthogonal to $J$).
A particular feature of the double scaling introduced in \cite{Hofman:2006xt} is that it decouples finite size effects of the system ($J< \infty$) from quantum effects ($\lambda \neq 0$), so it becomes possible to study the regime where the charges $J, \Delta$ are both finite.

For a closed spin chain in ${\cal N}=4$ SYM, the net central charge must vanish. This is related to the level matching constraint.
This condition can be relaxed if one considers open strings instead of closed strings.
More precisely, one can study giant magnon solutions for open strings that end on supersymmetric D-branes \cite{Berenstein:2013eya, Berenstein:2014isa}, rather than having an infinitely long spin chain.

 The central charge for the open string is sourced by the D-branes in a similar way than in the case of D-branes in flat space: the
physical separation of a pair of D-branes is the source of the central charge. In this case, the supersymmetric D-branes are giant gravitons \cite{McGreevy:2000cw} and the open strings between them are obtained by cutting the giant magnons for the sigma model at the location of the D-branes \cite{Berenstein:2014isa}. The results of the central extension match with traditional field theory computations up to various loop orders \cite{Dzienkowski:2015zba}.
These correspond to the central extension of the planar $\mathcal{N}=4$ SYM spin-chain studied by Beisert with open boundaries. Excitations of the central extended theory are expected to have an exact dispersion analogous to the above, given by
\begin{equation}
    \Delta- J = \sqrt{(k+1)^2+ \frac{\lambda}{4 \pi^2}|\xi - \tilde{\xi}|^2}\label{eq:disp}
\end{equation}
where $|\xi-\tilde{\xi}|$ is the Beisert central charge of the open string , and $k$ is a discretized angular momentum on the $S^5$. The variables $\xi, \tilde \xi$ are (complex) coordinate  positions of D-branes, and they live on a disk of radius one $|\xi|< 1$.
In the field theory dual, the coordinates $\xi$ appear naturally when addressing collective coordinate descriptions for giant graviton states \cite{Berenstein:2013md}. See also the appendix for details.
The boundary conditions on the spin chain follow from the works \cite{deMelloKoch:2007rqf,deMelloKoch:2007nbd,Bekker:2007ea} (see also \cite{Berenstein:2006qk}).

The relativistic form of \eqref{eq:disp} suggests the identification of  $|\xi-\tilde{\xi}|$  with the mass of the BPS string stretched between the branes; this is a quantity that is analogous to a length of a string.
Beisert's result is recovered from the open spin chain when we take the limit of the positions to the edge of the disk $|\xi|, |\tilde \xi| \to 1$.
These expressions have been also obtained from studying open strings on $\mathbb{R}_t\times S^n \subset AdS_5 \times S^5$ for both infinite strings and finite strings.

It is expected that similar expressions hold for the $SL(2)$ sector of the spin-chain with open boundaries.  In this case the central extension is actually the same central extension that determines the masses of W-bosons in the Coulomb branch of $\mathcal{N}=4$ SYM, as argued in \cite{Berenstein:2014zxa}. Recently, a combinatorial origin of this central extension in computations of anomalous dimensions in field theory has been explored in \cite{deMelloKoch:2020agz} (see also references therein). The corresponding analysis of the open $SL(2)$ spin chain and their boundary conditions is not understood yet.

The purpose of this paper is to find the classical solutions for the open strings that are associated to the $SL(2)$ sector.
 In this new case, the D-branes that source the central charge that correspond to the classical Coulomb branch of the field theory are the so called dual giant gravitons \cite{Grisaru:2000zn,Hashimoto:2000zp}. The Coulomb branch arises from classical solutions of the SYM equations of motion in flat space. Associated field theory configurations that describe the D-branes can be also treated semiclassically for the field theory on $S^3\times \BR$, as single eigenvalue solutions of the classical BPS equations of motion  for the dynamics (see  sections 3,4 of \cite{Berenstein:2007wi}).
 These expand into the $AdS$ directions, rather than on the sphere.
Because of this, the magnons, rather than looking like a point-like particle at the center of AdS, need to become extended in the AdS directions as well.
Naively, since $SO(4,2)$ has the same (complexified) Lie algebra as $SO(6)$, one expects that the corresponding classical strings should arise from an analytic continuation of
those solutions already found for the sigma model of $S^3\times \BR$, but now as solutions in $AdS_3\times \BR$. Our goal is to show how this analytic continuation works in detail for these solutions and to also explain the origin of the analogous variables $\xi, \tilde \xi$ for dual giant gravitons directly from the string sigma model.
Some  classical solutions that seem to address some of these issues, but that do not include the dual giant gravitons, have been studied in \cite{Ryang:2006yq}.
Also, the momentum $k$ rather than being an $SO(6)$ angular momentum, becomes an angular momentum on $AdS_5$.

These solutions will be called in this paper {\em open giant magnons}, which are suspended between dual giant gravitons. In the sigma model, these are  strings stretching between dual giant gravitons that extend in the $AdS_5$ directions. More precisely, we study such strings between dual giant gravitons that respect the same half of the supersymmetry.

We show in this paper that indeed, the dispersion relation of these giant magnons is given by a continous limit of \eqref{eq:disp}, where the central charge now corresponds to length of the open string stretched along the $AdS$ directions, and now $\xi, \tilde \xi$ will both be complex coordinates that are outside the disk of radius one. There is an additional constraint for the string solution that
depend on the precise position of $\xi, \tilde \xi$: the straight line from $\xi$ to $\tilde \xi$ in the complex plane must not cross the disk of radius one.
There are hints that this type of condition arises from perturbative calculations from studying the $SU(2)$ sector in \cite{Berenstein:2014zxa}, rather than the $SL(2)$ sector.
We will explain the origin of this constraint in this paper.

The paper is organized as follows.  In section \ref{sec:review} we review the solutions for giant magnons in the sigma model. We also interpret those solutions in terms of a geodesic on a disk inside the LLM plane of $AdS_5 \times S^5$, and their generalizations to two-charge magnons and we discuss their open string versions ending on giant gravitons.
 In section \ref{sec:openAdS}  we study classical strings on $AdS_2 \times S^1$ and $AdS_3 \times S^1$. These are a particular analytic continuation of the string solutions on $ \mathbb{R}\times S^n $ which we describe. A particular simplification is possible due to the boundary conditions necessary on the giant gravitons. The interpretation of these solutions in terms of the LLM plane is also discussed.

\section{Review of Giant Magnon Solutions}\label{sec:review}
Let us recall the scaling limit of the giant magnons of Hoffman and Maldacena \cite{Hofman:2006xt}. We are interested in the following scaling limit:

\beq
\begin{aligned}
J\;, \; \Delta &\rightarrow \infty \\
\lambda\;, \; p &< \infty\\
\Delta- J= \epsilon &< \infty
\end{aligned}
\eeq
Where $J$ is one of the $SO(6)$ R-charges, $p$ is the momentum of the excitation, and $\lambda= g^2 N$ is the t' Hooft coupling. In order for the semi-classical string description to be valid we should also consider the t' Hooft coupling $\lambda$ to be large. Then we seek for the solution with the least energy $\epsilon$ for a fixed momentum $p$.
The simplest of such configuration is given by a string that sits at the origin of $AdS_5$ while its endpoints rotate along the equator of $S^5$.  The motion takes place on $\mathbb{R} \times S^2$

\begin{equation}
    ds^2=-dt^2+ \cos^2 \psi d\theta^2+ d \psi^2.\label{eq:rtimessphere}
\end{equation}
Note that the spatial part of the metric  is that of a sphere of radius one with two coordinate singularities at $\psi= -\pi/2$ and $\pi /2$, where we have used a cosine of the angle, rather the sine of the angle. We have chosen a slightly different convention for the the metric of the sphere in order to make the analytic continuation into $AdS_2$ clear.  We can choose a parametrization for a rigidly rotating worldsheet coordinates for the Nambu-Goto string given by
\begin{equation}
\begin{aligned}
\tau&= t\\
\sigma& = \theta- t\\
\dot{\psi}&=0
\end{aligned}
\end{equation}
The condition $\dot \theta=1$ arises from the fact that the string becomes asymptotically a ferromagnet for the $SU(2)$ chain. In the presence of giant gravitons, this is the correct  coordinate velocity for the motion of the giant gravitons themselves.

Upon substitution  of the rigid ansatz, we get the action
\begin{equation}
    S= \frac{\sqrt{\lambda}}{2 \pi}\int d\tau d\sigma \sqrt{\sin^2 \psi \psi'^2+ \cos^2 \psi}.
    \label{eq:effecact}
\end{equation}
Using the coordinate transformation $r= \cos \psi $, minimizing the action \eqref{eq:effecact} takes the form of a simple geodesic problem with an effective  metric $ds^2= dr^2+ r^2d\theta^2$.
This is not the original metric of $AdS_5\times S^5$, but it is a flat auxiliary geometry. By virtue of $r\leq 1$, this is a flat metric on a disk.
 The conserved charges are given by
\beq
\begin{aligned}
\Delta & = \frac{\sqrt{\lambda}}{2\pi } \int d\sigma \frac{\psi'^2+ \cos^2 \psi }{\sqrt{\psi'^2 \sin ^2 \psi + \cos^2 \psi }}\\
 J &=  \frac{\sqrt{\lambda}}{2\pi } \int d\sigma \frac{\psi'^2\cos^2 \psi }{\sqrt{\psi'^2 \sin ^2 \psi + \cos^2 \psi }}
\end{aligned}
\eeq
It is convenient to write them also in terms of the $r$ variables:
\beq
\begin{aligned}
\Delta & = \frac{\sqrt{\lambda}}{2\pi } \int d\sigma \frac{(r^{\prime})^2(1-r^2)^{-1}+r^2}{\sqrt{r'^2 + r^2 }}\\
 J &=  \frac{\sqrt{\lambda}}{2\pi } \int d\sigma \frac{(r^{\prime})^2 r^2(1-r^2)^{-1}}{\sqrt{r'^2 + r^2 }}
\end{aligned}
\eeq
These develop  a singularity whenever $r\to 1$ in the solution. Notice that by contrast
\beq
\epsilon=\Delta-J= \frac{\sqrt{\lambda}}{2\pi } \int d\sigma {\sqrt{r'^2 + r^2 }}
\eeq
is non-singular and stays finite. More to the point, for these simple strings, extremizing  $\epsilon$ is the same as minimizing the geodesic problem we found above.

Note that $|r|\leq 1$ in order for the solutions to make sense, as $r=\cos\psi$ and the angle $\psi$ is a real coordinate of the sigma model. Near $r=1$ both of these expressions scale with $(1-r^2)^{-1}$. Substituting the explicit solution $r= a \sec \sigma $, the expression for $\psi'^2$ diverges whenever $a \sec \sigma =1$:
\beq
\psi'^2 = \frac{a^2 \sec^2\sigma\tan^2 \sigma }{1-a^2 \sec^2 \sigma}
\eeq
so the density of the conserved charges becomes infinite near such points. This is how one can have a smooth ending on an infinite spin chain that has not been closed.
However the effective energy of the configuration, $\Delta-J$ remains finite and it is the same as the on-shell action, which is clearly the length of a straight line segment connecting the two points on the edge of the auxiliary disk geometry:
\begin{equation}
\Delta- J= \frac{\sqrt{\lambda}}{ \pi} \Bigr|\sin \frac{\Delta\theta}{2}\Bigl|
\end{equation}
The angle between the two end-points is identified with the momentum of the defect $p$, as originally noticed in \cite{Berenstein:2005jq} by a matrix model ansatz. For closed string solutions we can sew together various of these straight line solutions on the disk such that in the end $\Delta \theta =0$, forming a closed polygon. This is equivalent to the level matching condition of a closed string $p_{total}=0$. \\

The case of open strings stretched between giant gravitons is entirely analogous except that one must impose that the endpoints of the string lie inside the disk.
 The end points must be
attached to the location of the giant gravitons.  These giant gravitons preserve the same half of the supersymmetry.
As the $S^3$ on which the branes are wrapped  shrinks to zero size at $r=1$, these do not exit the disk. The analysis has been carried out in \cite{Berenstein:2014isa}.

As a result of the end-points not reaching $r=1$, both of the charges $\Delta, J$ remain finite. The disk coordinates $ds^2= dr^2+r^2 d\psi^2$ can then be  seen to be the coordinates for the LLM plane of $AdS_5 \times S^5$ \cite{Lin:2004nb}. This description of the physics in terms of  a disk also appears directly from the field theory dual \cite{Berenstein:2004kk}.

\section{ Open Giant Magnons in AdS}\label{sec:openAdS}
\subsection{$AdS_2 \times S^1$}
Now we proceed to study the analogous solutions for the case of open strings stretching between two D3 branes that wrap an $S^3$ inside $AdS_5$ while rotating at angular velocity $\omega =1$ along an $S^1 \subset S^5$. As in the previous section we may consider the Nambu-Goto action for a string on an $AdS_2\times S^1$ geometry. The metric is given by
\begin{equation}
ds^2 = - \cosh^2\rho dt^2 + d\rho^2 +d\phi^2.\label{eq:ads2timesr}
\end{equation}
We will be interested in solutions of the equations of motion where the string rigidly rotates with the dual giant gravitons.
As such, $\phi$ will evolve in time in the same way that the coordinates of the dual giants do. These rotate  at constant velocity in $\phi$, have a fixed value of $\phi$ at each time and are located at fixed values of $\rho$  \cite{Hashimoto:2000zp, Grisaru:2000zn}.
That is, $\dot \phi=1$. These are a different  type of solution to the rigidly rotating  GKP string \cite{Gubser:2002tv}, as they have motion in one extra dimension.

We choose to parametrize the worldsheet coordinates by $t=\tau$, $\phi= \tau+ \sigma$ and we will be looking for solutions where $\rho(\sigma)$ is independent of $\tau$.
This condition with $\dot\phi=1$ has to do with the motion of the dual giant gravitons that move at speed one in the $\phi$ direction.
To have a reduction to $AdS_2\times S^1$, the two dual giants must preserve the same half of the supersymmetry.

The induced metric on the string worldsheet for these solutions is given by
\begin{equation}
ds^2_{ind} = \begin{pmatrix}
-\cosh^2 \rho +1 & 1\\
1& 1+(\rho')^2.
\end{pmatrix}
\end{equation}

This way we find that
\begin{equation}
S= -\frac{\sqrt{\lambda}}{2 \pi} \int d\tau d\sigma \sqrt{-g}=- \frac{\sqrt{\lambda}}{2 \pi}\int d\tau d\sigma  \sqrt{\sinh^2 (\rho) (\rho')^2+\cosh^2 (\rho)} \label{eq:effecact2}
\end{equation}
which can be seen to be an analytic continuation of the sigma model action of the solutions of the giant magnons of
Hoffman and Maldacena, $\psi \rightarrow i \rho$. The problem again simplifies by introducing the change of variables $r = \cosh(\rho)$, so that
\begin{equation}
S \propto \tau \int d\sigma \sqrt{r'^2 +r^2 }
\end{equation}
The expression $\int d\sigma \sqrt{r'^2 +r^2 }$ can be easily seen to be the length of a curve on a flat geometry in polar coordinates $d\tilde s^2= dr^2+ r^2 d\sigma^2$, in
a parametrization $r(\sigma)$. This is minimized by a straight line, where
\begin{equation}
r = \frac a{\cos(\sigma-\sigma_0)}
\end{equation}
where $a$ is the distance of closest approach to the origin and $\sigma_0$ is the angle in the plane of closest approach.
The energy computed this way is also proportional to the length of the straight line in this auxiliary geometry from the starting point to the end point.

It is important to note that although superficially similar to the solutions of \eqref{eq:effecact}, for the change of variables  to make sense in this case we must also impose that $|r|> 1$ everywhere on the worldsheet, the reason being the $S^3\subset AdS^5$ shrinks to zero size at $r=1$. In particular, solutions to \eqref{eq:effecact2} which cross the unit circle are not physical, as they would require the radial coordinate of $AdS$ to become complex. Even though there are in principle solutions of minimal length when one removes the inside of the disk, these are not stable in that they should receive quantum corrections since they are no longer BPS.

The coordinates of this auxiliary plane geometry span the region outside the disk in the LLM plane, rather than the inside. Solutions with $r=1$ somewhere have a similar behavior to the Hofmann Maldacena solution, in that the density of the charges $\Delta, J$ becomes infinite at $r=1$, yet the energy $\Delta-J$ remains finite.

The introduction of a third charge is straightforward. We will do that analysis in the discussion of the following section. We mostly follow the discussion in \cite{Minahan:2006bd}, which starts from the Nambu-Goto string, to make the analysis.

 \subsection{Rotating String in $AdS_3\times S^1$}
 Now we consider the sigma model of a rotating string on $AdS_3 \times S^1$ which corresponds to a two-spin magnon solution. The metric is a simple generalization of \eqref{eq:ads2timesr}

 \begin{equation}
     ds^2= - \cosh^2 \rho dt^2 + d \rho^2+ \sinh^2 \rho  d \theta^2 + d\phi^2\label{eq:ads3timesr}
 \end{equation}
Here again it is convenient to use the coordinate $r= \cosh \rho$ in the Nambu-Goto string action. As noticed in the appendix, the coordinate $r=\cosh(\rho)$ is  the radius of the LLM plane coordinate.

The metric \eqref{eq:ads3timesr}
can be analytically continued via $\rho \rightarrow i \tilde{\rho}$ into $\mathbb{R}\times S^3$ where the $S^3$ is expressed in Hopf coordinates.
 We make an ansatz for the embedding coordinates of the form :

 \begin{equation}
     \begin{aligned}
     t &= \omega_t \tau\\
     r &= r(\sigma)\\
     \theta &= \beta \tau+ g(\sigma)\\
      \phi&= \omega_{\phi} \tau + \varphi(\sigma)
     \end{aligned}
 \end{equation}
 Where we are interested in $\omega_t= \omega_{\phi}=1$.

 The action for the rigid string in these coordinates is given by:

 \begin{equation}
     S = -\frac{\sqrt{\lambda}}{2 \pi} \int d\tau d\sigma \sqrt{r'^2+ \varphi'^2 + 2 \beta g' \varphi'(r^2-1)+g'^2(r^2-1)^2-\beta^2(r'^2+ \varphi'^2(r^2-1)) },
 \end{equation}
 where we have set the angular frequencies to one for simplicity. The conserved quantities can be easily evaluated via the formulas:
 \begin{equation}
    \begin{aligned}
    \Delta &= -\int d\sigma \frac{\partial \mathcal{L}}{\partial \omega_t}\Bigr|_{\omega_t =1}\\
    J_1 &= \int d\sigma \frac{\partial \mathcal{L}}{\partial \omega_\phi}\Bigr|_{\omega_\phi =1}\\
    J_2 &= \int d\sigma \frac{\partial \mathcal{L}}{\partial \beta}\\
    \end{aligned}
 \end{equation}
As we will see, one can eliminate the angular variable $g$ by using its equation of motion:
 \begin{equation}
     \partial_{\sigma}\left( \frac{(r^2-1)(\beta  \varphi'+ (r^2-1) g') }{\sqrt{r'^2+ \varphi'^2 + 2 \beta g' \varphi'(r^2-1)+g'^2(r^2-1)^2-\beta^2(r'^2+ \varphi'^2(r^2-1)) }} \right)= \partial_\sigma {\mathfrak J}^\sigma=  0 \label{eq:eomg}
 \end{equation}
 We have chosen to write it in terms of an expression that is implied by a current conservation, which forces ${\mathfrak J}^\sigma$ to be constant.
At first solving for $g$ might seem daunting as \eqref{eq:eomg} reduces to a complicated equation depending on an integrating constant, which is the value of $ {\mathfrak J}^\sigma$. A great simplification is possible since in the end we are interested in describing strings ending on a pair of giant gravitons, so one must be careful about imposing the correct boundary conditions at the string end points.

The  boundary term that arises from taking the variation of the action (3.8) is of the form:
 \begin{equation}
     S_{bdy} \propto \int d\tau \left(\delta \theta \;  {\mathfrak J}^\sigma\right)\Bigr|^{\sigma_f}_{\sigma_i}
 \end{equation}
 Where ${\mathfrak J}^\sigma= \frac{\partial \mathcal{L}}{\partial\left(\partial_\sigma \theta \right)}$ is precisely the quantity inside the parentheses of (3.10). The other boundary terms are set to vanish by imposing the appropriate Dirichlet boundary conditions $\delta r= \delta \phi=0$ and by the choice of static gauge $\partial_\sigma t=0$.
 Because the D-brane we are considering is extended in the $\theta$ direction, the correct boundary conditions for $\theta$ are Neumann boundary conditions, which means that $\delta \theta$ is free to vary.

 We must then conclude that ${\mathfrak J}^\sigma$ vanishes at the end-points of the string in order for the variational principle to be well-defined. In addition to this, equation
 \eqref{eq:eomg} implies that this quantity vanishes everywhere along the string. This allows us to solve implicitly for the function $g$ in terms of the other coordinates:
 \begin{equation}
     \partial_\sigma g= - \beta\frac{ \partial_\sigma \varphi}{r^2-1}
 \end{equation}
Solutions of this type have been previously considered for infinite strings in the $S^3 \times \BR$ sigma model (see \cite{Minahan:2006bd,Ryang:2006yq}), but their physical interpretation was not made clear.
Here the meaning of the condition is clear: the giant magnon does not transport angular momentum in the $\theta$ direction from one D-brane to the other. The giant magnon carries that angular momentum, but it does not transfer it to the D-branes.
 Eliminating the variable $g$ in this case does not affect the variational principle for $r$, so one may substitute  that  condition directly
 in order to express the conserved quantities in terms of the on-shell action multiplied by some kinematic factors:

 \begin{equation}
 \begin{aligned}
 \Delta- J &= \frac{\sqrt{\lambda}}{2\pi} \frac{\mathcal{Z}}{\sqrt{1-\beta^2}} \\
J_1 &=  \frac{\sqrt{\lambda}}{2\pi}\frac{\beta \; \mathcal{Z}}{\sqrt{1-\beta^2}} \\
 \mathcal{Z}&=-  \int_{\varphi_i} ^{\varphi_f} d\varphi\sqrt{r^2+ (\frac{dr}{d \varphi})^2}
\end{aligned}
 \end{equation}
 To do the variation, we want to minimize the energy, $\epsilon=\Delta -J$ at fixed $\beta$, with the endpoints on the dual giant gravitons.
 This is a straightforward minimization of $\mathcal{Z}$ that results in a straight line.
Clearly the variable $\mathcal{Z}$ corresponds (up to a factor) to the length of the string on an auxiliary flat 2D geometry with a disk removed, and as such should be identified with the central charge of the $SL(2)$ sector of the spin chain. The complex coordinate $\xi = r\exp(i\varphi)$ can be used to express the answer in terms
 similar to those of \eqref{eq:disp}. Here we see that $|\xi_1-\xi_2|$ is the length of the segment in the LLM plane connecting the two giant gravitons.

 Eliminating $\beta$ in terms of $J_2$ and $|\mathcal{Z}|$ yields the dispersion relation

 \begin{equation}
     \Delta- J_1= \sqrt{J_2^2+ \frac{\lambda}{4\pi^2}|\mathcal{Z}|^2}
 \end{equation}
 Which is precisely of the form expected from equation \eqref{eq:disp}. Although the boundary conditions considered here lead to solvable equations of motion for the ground state of the sigma model, it is not expected that these solutions (and their many magnon counterparts) can be fully described by an integrable spin chain. The boundary effects sourced by the branes are expected to  destroy that property. This has been argued, at least from the notion of a simple Bethe Ansatz point of view, for the open spin chains attached to regular giant gravitons in  \cite{Berenstein:2006qk}. It would be very interesting if a class of solutions of the sigma model of this type do lead to integrable boundary conditions for the dual spin chain description.

 It is also instructive to give explicit expressions for the conserved quantities $\Delta, J_1$. Since we are considering open strings of finite size, one would expect that these quantities are finite.
 However, for similar reasoning to that of the previous section, one must be careful that the radial coordinate doesn't touch the unit circle given by $r=1$. We can see this from the expressions:
 \begin{equation}
\begin{aligned}
 \Delta& = \frac{\sqrt{\lambda}}{2\pi}\int d\phi \frac{r^2(\frac{dr}{d \varphi})^2+ \beta^2-1+ r^4}{(r^2-1)\sqrt{(1-\beta^2)(r^2+ (\frac{dr}{d \varphi})^2)}}\\
J_1&=\frac{\sqrt{\lambda}}{2\pi}\int d\phi \frac{(\frac{dr}{d \varphi})^2+ \beta^2-1+ r^2}{(r^2-1)\sqrt{(1-\beta^2)(r^2+ (\frac{dr}{d \varphi})^2)}}
 \end{aligned}
 \end{equation}
 It's clear that these quantities diverge whenever $r=1$ (that is, if the string touches the origin of AdS), even though the length of the string on the auxiliary  plane geometry is finite as in the Hoffman-Maldacena string. Solutions like these, where a physical worldsheet quantity is becoming divergent should become unphysical and lead  to non-normalizable states, in a way similar to the discussion  in \cite{Berenstein:2014zxa}. More concretely, these lead to operators that would inject an infinite amount of energy into the bulk and where the radial coordinate becomes complex.

 \section{Discussion}
In this paper  we have provided evidence for the all-loop dispersion relation for the excitations of the $SL(2)$ sector of the  $\mathcal{N}=4$ SYM spin chain with open boundary conditions. The calculation is done in the sigma model and it gives rise to a series expansion in the t'Hooft coupling $\lambda$.
We find a nice description of the solutions in terms of an analytic continuation of the $SU(2)$ sigma model and the solution has a form that suggests that these states arise from a BPS
condition on the string worldsheet, as would be expected from shortening conditions of the central extension of the ${\cal N}=4 $ SYM spin chain \cite{Beisert:2005tm}.
It would be nice to understand this better from the planar $SL(2)$ sector of the ${\cal N}=4$ SYM theory in more detail.
This spin chain should precisely realize the open string sigma model in a continuum limit.  It would also be of interest to consider more general solutions to the string sigma model corresponding to scattering and bound states of giant magnons. After all, the open strings suspended between ordinary giant gravitons  have a relation to the Bethe ansatz of the $SU(2)$ spin chain, as noted in  \cite{Berenstein:2014isa}.

The description in the $SL(2)$ sector is expected to be  qualitatively different than the $SU(2)$ magnons due to the nontrivial boundary condition imposed by $|r|>1$.
This is already clear from the sigma model description, as constructing solutions from inverse scattering methods for the $SL(2)$ model seems to require different Bethe roots than the $SU(2)$ sigma model, even though their solutions should be related by an analytic continuation of the solutions in the $SU(2)$ sector \cite{Spradlin:2006wk}. It is unclear what the precise structure of bound and scattering states is without further study of the field theory side.

We also found that the LLM coordinates arise naturally from studies of the string sigma model in the $SL(2)$ sector.  This might allow us to understand better how certain aspects of locality in the radial direction of $AdS$ arise from the dual field theory directly. One should also expect that given these solutions, that one could also compute the spectrum of quadratic fluctuations around these solutions and could in principle compare to the ${\cal N}=4 $ SYM spin chain . This analysis can also be extended to general $\frac{1}{2}$ BPS geometries, where the states dual to closed string magnons have been studied \cite{Koch:2016jnm}. It is expected that the charges of open strings stretched between giants generically diverge as the branes approach the regions where their corresponding 3-spheres shrink to zero size. This suggests that one should expect instabilities that are not captured by the naive perturbative expansion of the gauge theory, and that one should make sure that the corresponding gauge theory operators can be matched with valid open string solutions on the LLM plane.

It would also be interesting to understand how such strings suspended between dual giant gravitons that preserve different halves of the supersymmetry would work out. This would allow us to better explore how the physics of the Coulomb branch of ${\cal N}=4 $ SYM determines the dynamics of the theory in more detail. The analogous problem with giant gravitons that preserve different supersymmetries has not been studied either.

 Similar simplifications as the ones found here should also be possible on backgrounds of the form $AdS_5 \times X_5$ where $X_5$ is a Sasaki-Einstein manifold, as they share the $AdS$ part of the sigma model, and the Sasaki-Einstein geometries have a $U(1)$ fibration that should allow solutions of a similar kind.  It would also be of use to study the case of $AdS_4 \times \mathbb{C}P^3$ with fluxes, related to the ABJM model \cite{Aharony:2008ug}. The latter should be quite interesting, as the exact characteristic of the allowed brane configurations depend greatly on the details of the field theory set-up.

\acknowledgments

The work of D.B. is supported in part by the Department of Energy under grant DE-SC 0011702.

\appendix

\section{Giant gravitons and LLM Geometries}

The half BPS states of ${\cal N}=4$ SYM can be described in terms of a quantum hall droplet picture in two dimensions, with a
confining quadratic potential \cite{Berenstein:2004kk}. The droplet is a condensate of eigenvalues of a complex chiral scalar  matrix field $Z$, which is one of the 3 chiral multiplets of the ${\cal N}=4$ SYM theory in an ${\cal N}=1$ superfield notation.
The ground state of the field theory is described by a circular droplet.

In this picture giant gravitons are holes (vacancies) in the droplet, and dual giant gravitons are particles that are away from the droplet.
If the circular droplet has radius $R$, an eigenvalue at the edge of the droplet will have energy $\kappa R^2$. A vacancy in the center of the droplet is a maximal giant graviton with energy $N$, so that
$\kappa R^2 = N$, which is the amount of energy that the eigenvalue at the center gets when moved to the edge. For any other radius, a hole will have energy
$E=L = \kappa(R^2-r^2)$, where $E,L$ are the energy and angular momentum of the giant graviton (they are the same because of the BPS condition). This allows us to set up a correspondence between coordinates of the droplet plane and the energies of giant gravitons.
By comparison with the work \cite{McGreevy:2000cw}, we find that the giant graviton is at a radius
\begin{equation}
\frac rR = \sqrt{1-\frac{L}{N}}
\end{equation}
One can check that the radius $r$ together with a phase $\theta$ corresponds to a coordinate on a disk $\xi= r/R \exp(i \theta)$ that results from writing a five sphere as $\sum x_i^2=1$, and
considering $\xi = x_1+i x_2$ (see \cite{Berenstein:2014isa} for more details).

For dual giant gravitons we can use the same intuition. Now, a dual giant at radius $r$ will have energy given by
\begin{equation}
E= L=  \kappa (r^2-R^2)
\end{equation}
so that
\begin{equation}
\frac rR = \sqrt{1+\frac LN}
\end{equation}
In the field theory dual,  these are given by certain Young Tableaux states characterized by long rows \cite{Corley:2001zk}.
For global coordinates, where
\begin{equation}
ds^2 = -\cosh^2(\rho) dt^2 +d\rho^2 +\sinh^2(\rho) d\Omega_3^2
  \end{equation}
  a giant graviton will be at a radius of the three sphere characterized by
(see the discussion in \cite{Hashimoto:2000zp})
\begin{equation}
\sinh \rho = \sqrt{\frac LN}
\end{equation}
This is equivalent to
\begin{equation}
r= \cosh(\rho)= \sqrt{1+\frac LN}
\end{equation}
This identifies the coordinate $r$ of the droplet with $\cosh(\rho)$ in the bulk.
Indeed, the coordinates that arise this way are the LLM coordinates \cite{Lin:2004nb}.
The LLM supergravity solutions are given also by droplets on a plane, which can be completed to a ten dimensions geometry.
The LLM plane is the degeneration locus of two types of spheres. One of those arises from  the $S^5$ geometry and the other one from the  $AdS_5$ geometry.
 These are a complete classification of half BPS states
in the $AdS_5\times S^5$ geometry.

The excess energy of the solutions relative to the ground state is given by an integral over the droplet area
\begin{equation}
E\propto  \int_{\cal D} d^2 x (x_1^2+x_2^2) -  \int_{{\cal D}_0} d^2 x (x_1^2+x_2^2)
\end{equation}
where we do an integral over the area ${\cal D}$ that is covered by particles, where ${\cal D}_0$ is the ground state droplet.
We also have that $N\propto \int_{\cal D} d^2 x$ is the area covered by the droplet, and it is independent of the shape of the droplet. A single giant graviton
has a fixed unit of area.
We see that in the LLM geometry
the energy of a particle removed from the droplet is also quadratic in the coordinates of the LLM plane.

\end{document}